\documentclass{ws-p8-50x6-00}

\newcommand{\rp}{\mbox{$\not \hspace{-0.15cm} R_p$} }

\begin{document}

\title{Searches at HERA}

\author{Nicolas Delerue\\(on behalf of the H1 and ZEUS collaborations)}

\address{Centre de Physique des Particules de Marseille\\
163, avenue de Luminy,\\
Case 907,\\
F-13288 Marseille Cedex 9\\
France\\
E-mail: \tt nicolas.delerue@desy.de}

\maketitle

\abstracts{Searches for physics beyond the Standard Model have been performed in high-energy $e^{\pm}p$ collisions at HERA. No significant deviation from the Standard Model has been observed while searching for contact interactions, extra dimensions, leptoquarks, R-parity violating squarks and excited fermions. Exclusion limits have been inferred which extend or complement bounds from other colliders. The H1 collaboration has observed a puzzling excess of events with an high $P_t$ isolated lepton and missing transverse momentum, and interpretation as flavour changing neutral currents has been explored.}

\section{Introduction}
 
\subsection{HERA}
HERA is the first and only existing lepton-proton collider. Between 1994 and 2000 it has delivered an integrated luminosity of more than 130 pb$^{-1}$ available for physics to each experiment. Leptons (electrons or positrons) have an energy of 27.6 GeV and protons have an energy of 920 GeV (or 820~GeV), this gives an energy in the centre of mass ($\sqrt{s}$) of 318 GeV (or 300~GeV). 
The HERA collider is now shut down for a luminosity upgrade.

\subsection{Physics at HERA}


One of the major Standard Model (SM) physics processes at HERA is  deep inelastic scattering (DIS) where the incoming lepton exchanges a gauge boson with a quark of the proton. 
Neutral Current (NC) events result from the exchange of a neutral ($\gamma$,$Z^0$) boson and yield an electron and a recoil jet in the final state, Charged Current (CC) events result from the exchange of a charged  ($W^{\pm}$) boson and yield a neutrino and a recoil jet in the final state.



\section{Inclusive NC DIS events at high $Q^2$}

Phenomena characterised by a mass scale $\Lambda$ larger than HERA's kinematic limit ($\sqrt{s} \simeq 318 $ GeV) can be parameterised by four-fermion interactions which could affect the cross-sections for DIS observed at HERA via contact interactions. Comparisons of the cross-section predicted by the SM and those observed at high $Q^2$ at HERA have been made by both collaborations~\cite{Adloff:2000dp,Breitweg:2000ssa}. No deviations have been observed for $Q^2$ up to 10,000 GeV$^2$. Thus, exclusion limits in various contact interaction models can be set on $\Lambda$ up to 9 GeV with the usual coupling strength convention of $\sqrt{4\pi}$. 
Limits on large extra dimensions have also been set by H1 and exclude scale parameters ($M_s$) lower than 0.63 TeV and 0.93 TeV for positive and negative interference, respectively.

\section{Leptoquarks}

Leptoquarks are scalar or vector bosons  which couple to a quark and a lepton via a Yukawa coupling $\lambda$. They can be resonantly produced at HERA.  Their final topology  is the same as NC or CC events but their angular distribution can be used to optimise the signal significance.

\begin{figure}[htbp]
\begin{center}
\begin{tabular}{cc}
\epsfig{file=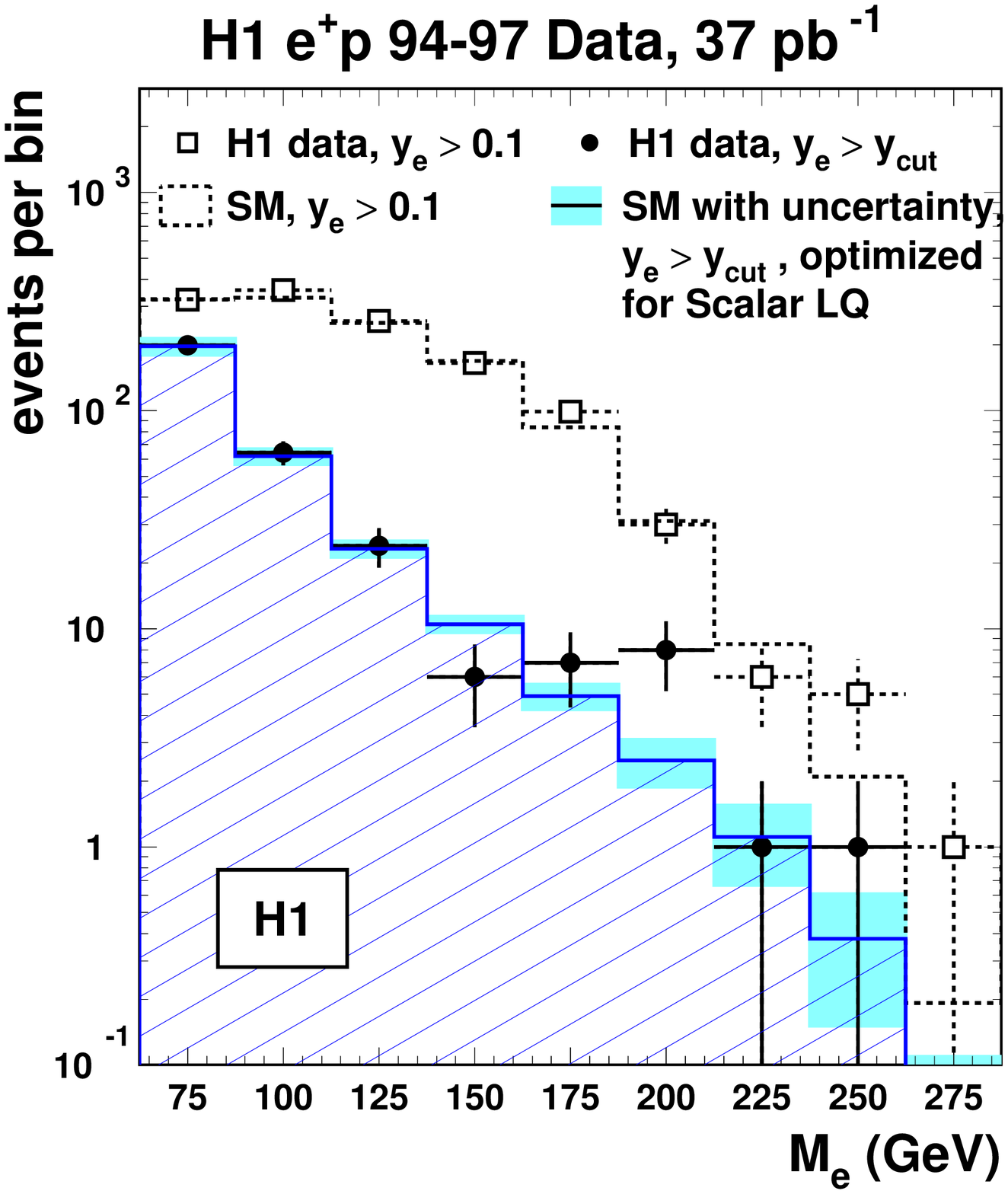,width=4.cm}&\epsfig{file=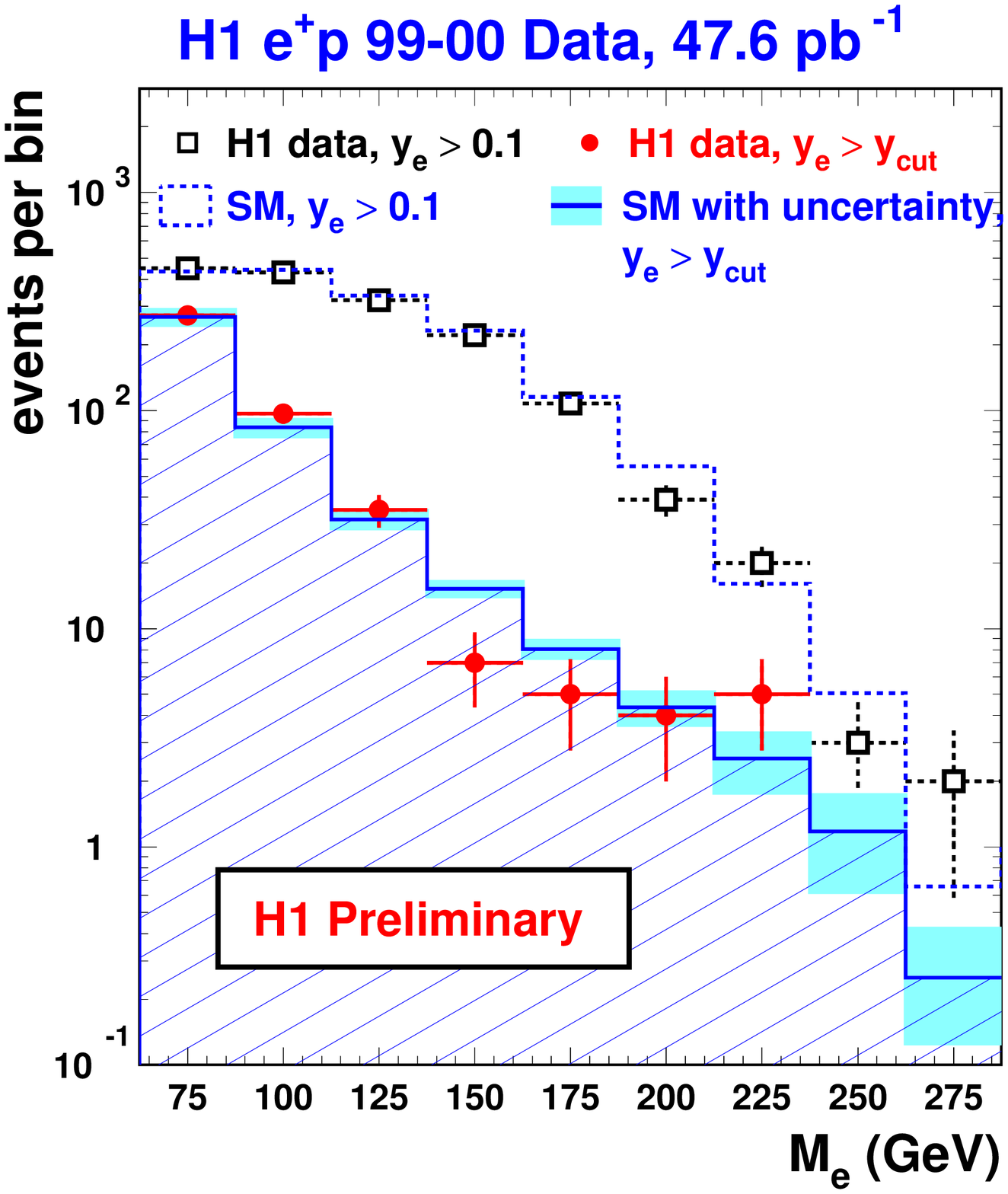,width=4.cm}
\end{tabular}
\end{center}
\vspace*{-.5cm}
\caption{The mass distribution of observed NC-DIS events compared to the  expected background for the 1994-1997 data (left) and 1999-2000 data (right).\label{fig:lqMassPlot}}
\vspace*{-.5cm}
\end{figure}

It had been shown earlier~\cite{Adloff:1999tp,Breitweg:2000sa} that the HERA data collected in 1994-97 showed an excess of electron-quark masses greater than 200 GeV (figure~\ref{fig:lqMassPlot}, left). This excess is not confirmed in the new data collected by both experiments\cite{osk-953,osk-1038,osk-954} (figure~\ref{fig:lqMassPlot}, right).

\begin{figure}[htbp]
\begin{center}
\begin{tabular}{c}
\epsfig{file=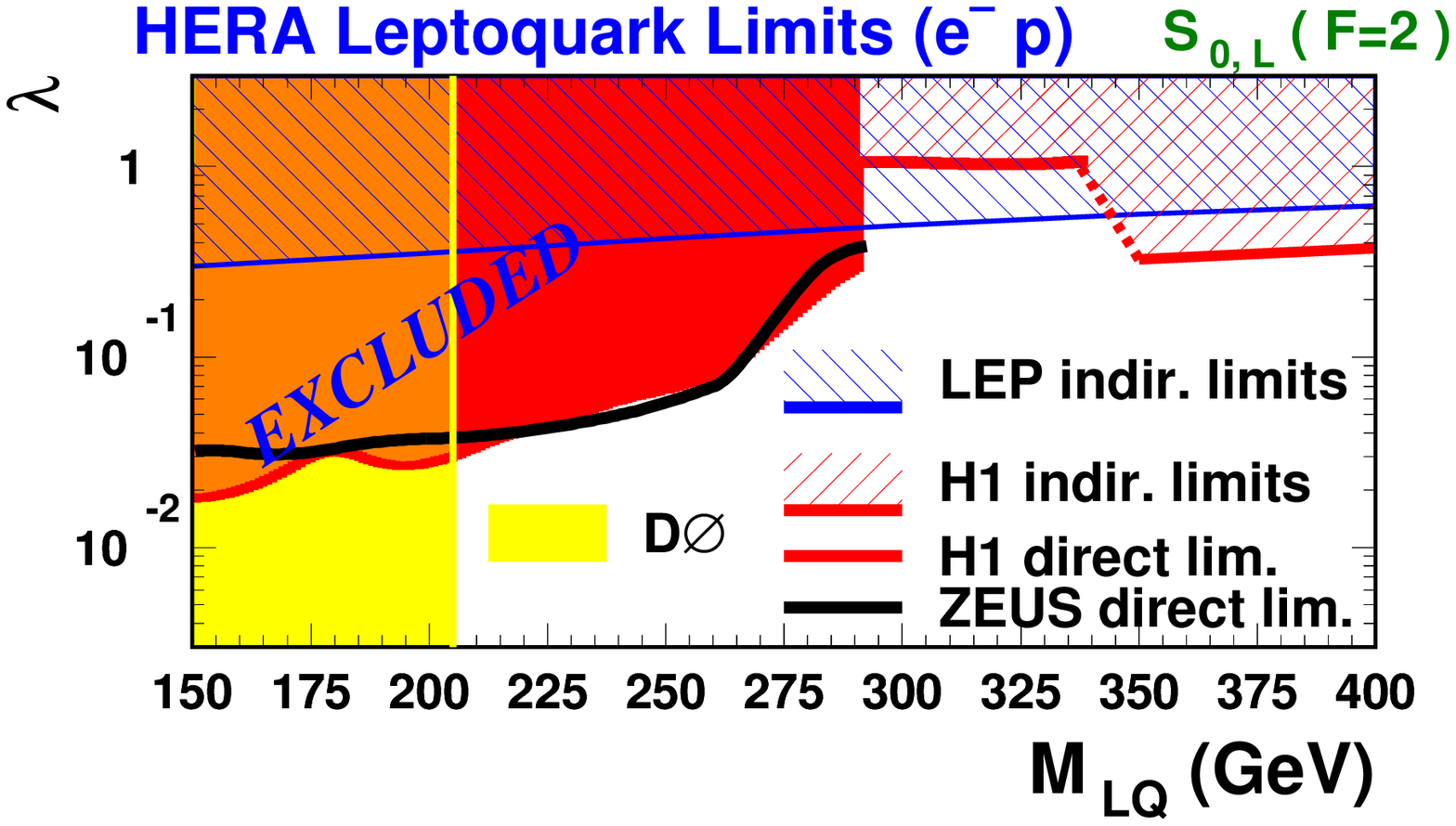,width=7.5cm}
\vspace{2cm}\\
\hspace*{-2cm}\epsfig{file=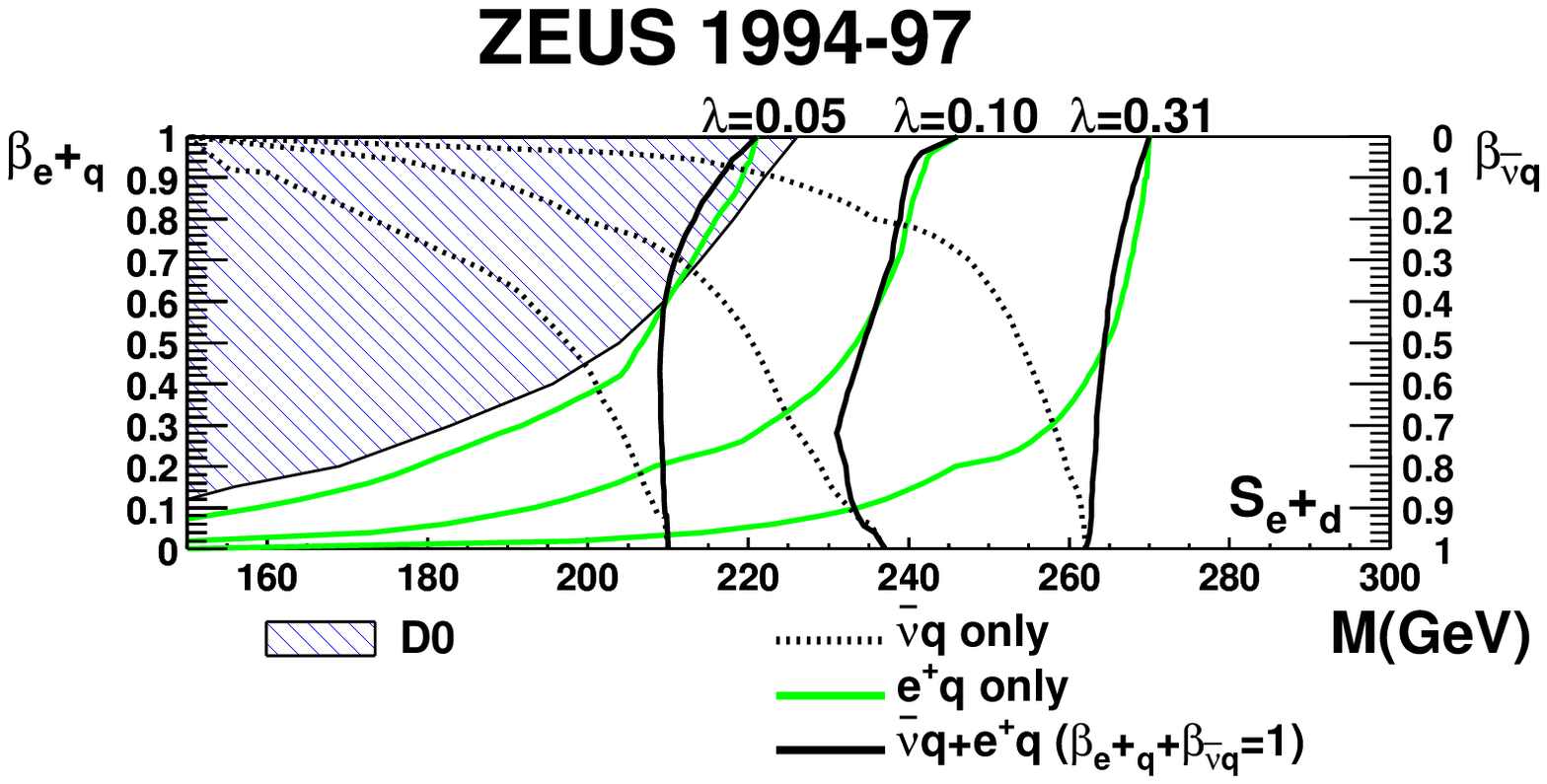,width=8.cm}
\end{tabular}
\end{center}
\vspace*{2cm}
\caption{\label{fig:lqLimPlot} Limits on the Yukawa coupling in the BRW model (upper plot) and limits on the branching ratio $ LQ \rightarrow eq $ versus mass for fixed values of $\lambda$ (lower plot).}
\vspace*{-.5cm}
\end{figure}

As no excess is seen, limits have been derived by both collaborations. The results have been interpreted in the model of Buchm\"uller, R\"uckl and Wyler (BRW)~\cite{Buchmuller:1987zs} where the leptoquark branching ratios are fixed.
 The upper plot of figure~\ref{fig:lqLimPlot} shows the limits set by both collaborations in the BRW framework. The lower plot shows the limits derived in a more general model for fixed values of the coupling\cite{Breitweg:2001ks}.

\section{\rp supersymmetry}

$R_p$-violating squarks can be produced at HERA up to the kinematic limit in the same way as leptoquarks. They would have many possible decay modes and many different topologies, including some very striking, almost background-free, topologies such as events with a lepton whose charge is opposite to that one of the incoming lepton, or with multileptons.

\begin{figure}[htbp]
\begin{center}
\begin{tabular}{cc}
\epsfig{file=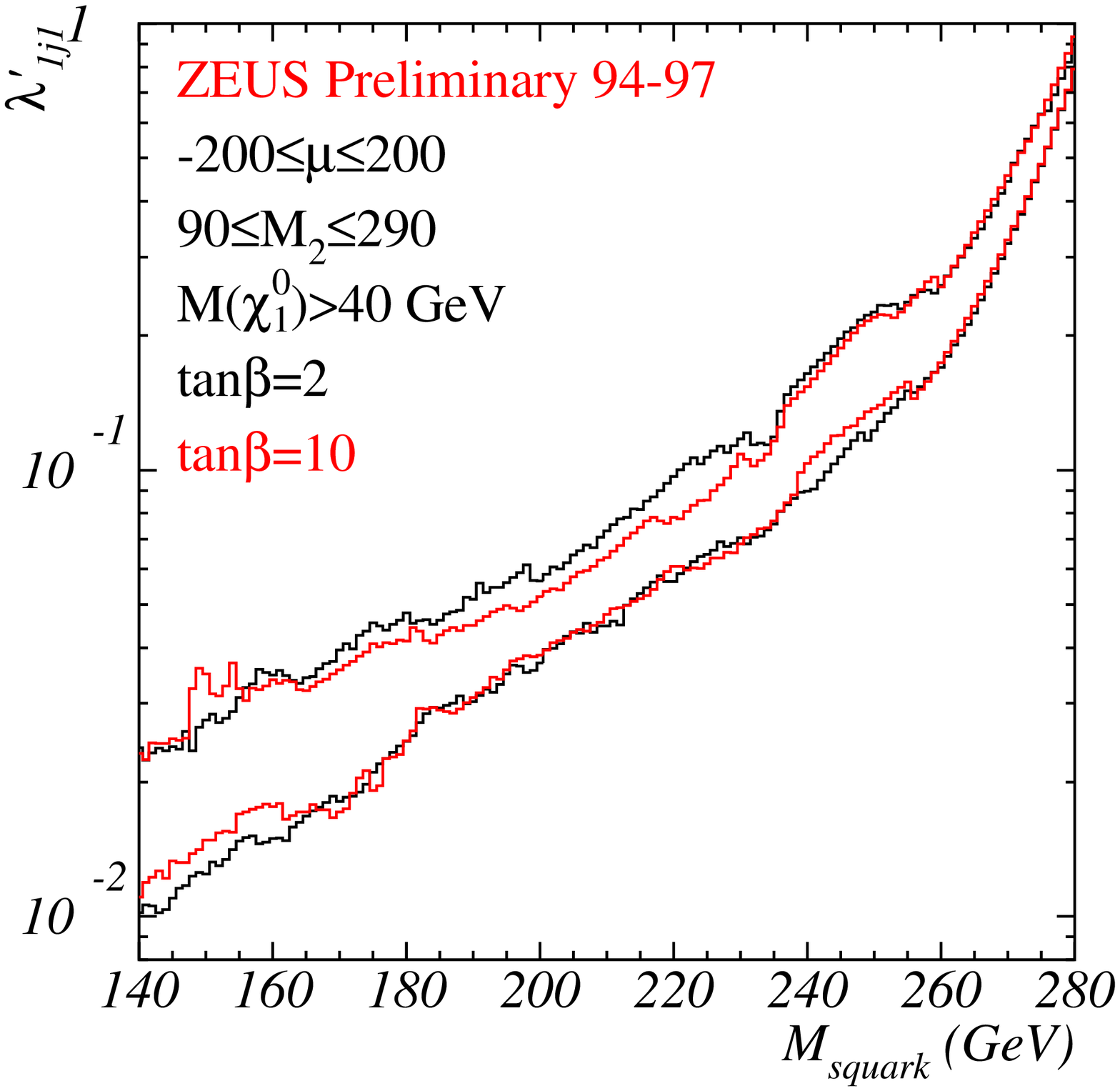,width=6.cm}&
\epsfig{file=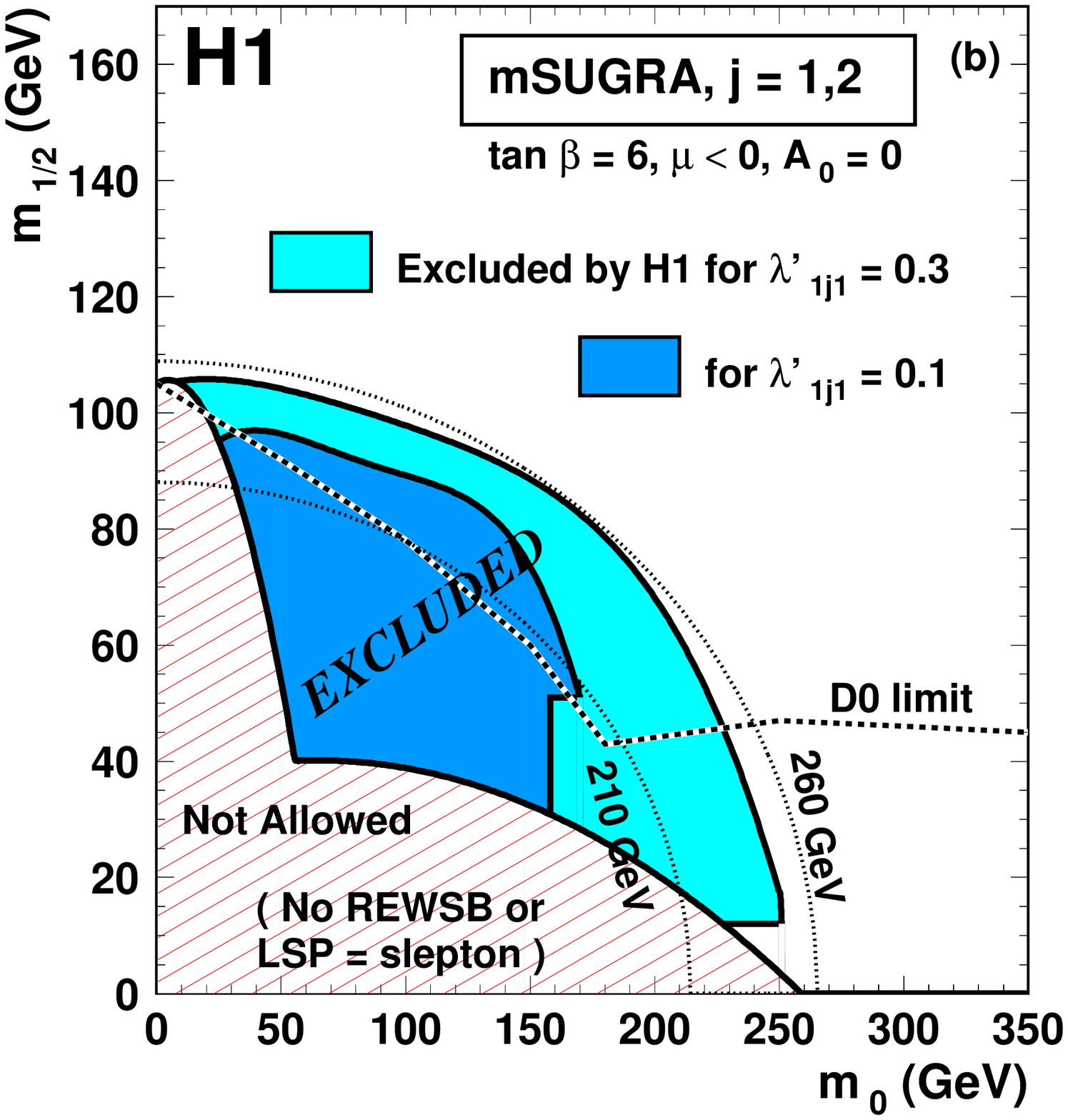,width=6.cm}\\
\end{tabular}
\end{center}
\vspace*{-.5cm}
\caption{{\it Left:} Limits on $\lambda'_{1j1}$ as a function of the squark mass for the unconstrained MSSM. For each $\tan \beta$ hypothesis (dark lines for $\tan \beta = 2$ and light gray lines for $\tan \beta = 10$), the best and worse limit found during the scan are shown
{\it Right:} Domain of the plane ($m_0,m_{1/2}$) excluded by the H1 analysis for $\mu < 0$ and $\tan \beta = 6$ for a \rp coupling of $\lambda'_{1j1} = 0.3 $ (light shaded area) and  $\lambda'_{1j1} = 0.1 $ (dark grey area). The region below the dashed curve corresponds to the area excluded by the D0 experiment.\label{fig:SUSY}}
\vspace*{-.5cm}
\end{figure}

The signal was looked for in almost all decay modes and no excess was observed~\cite{Adloff:2001at,osk-1042}. Upper limits on the Yukawa coupling $\lambda_{1j1}'$ have been obtained as a function
    of the squark mass (see figure~\ref{fig:SUSY}, left), by performing
    a scan in the parameters of the Minimal
    Supersymmetric Standard Model (MSSM). For a coupling 
    of the electromagnetic strength, squark masses below 258 GeV
are ruled out.
    Constraints on the parameters of the
mSUGRA
 model are also shown in figure~\ref{fig:SUSY} (right) for
     specific values of the Yukawa coupling.

\section{Excited fermions}

Compositeness could manifest in excited states of fermions. The single production of excited fermions is considered at HERA. An effective Lagrangian~\cite{Kuhn:1984rj,Hagiwara:1985wt} describes the production and decay of excited fermions. In this Lagrangian, $\Lambda$~is the compositeness scale and $f$, $f'$ and $f_s$ are form factors which determine the couplings of a fermion to its corresponding excited states and to the bosons associated to the three gauge groups (SU(2), U(1) and SU(3)).

%

\begin{figure}[htbp]
\begin{center}
\vspace*{-2.3cm}
\begin{tabular}{ccc}
\hspace*{1cm}\epsfig{file=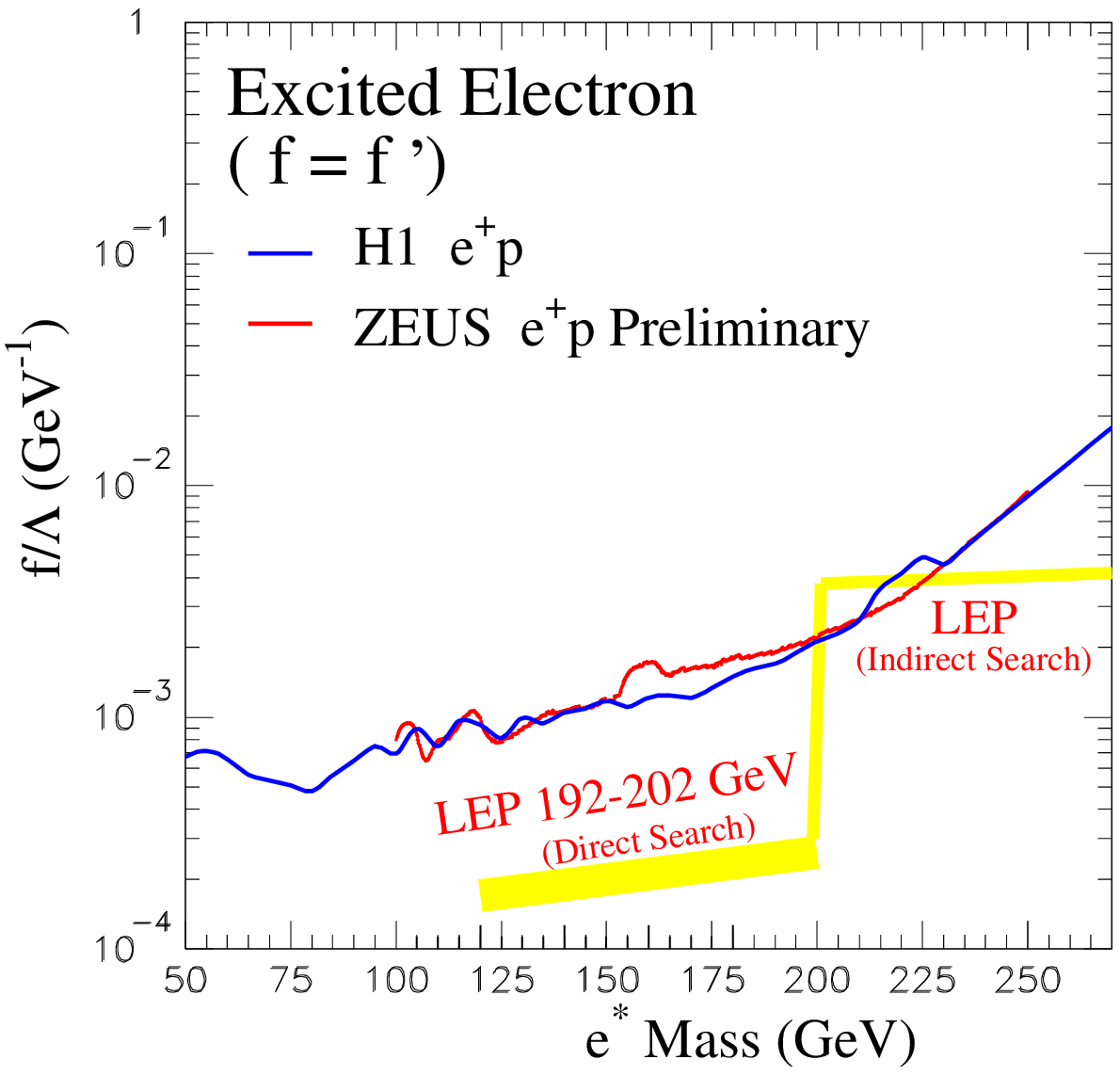,width=3.7cm}&
\epsfig{file=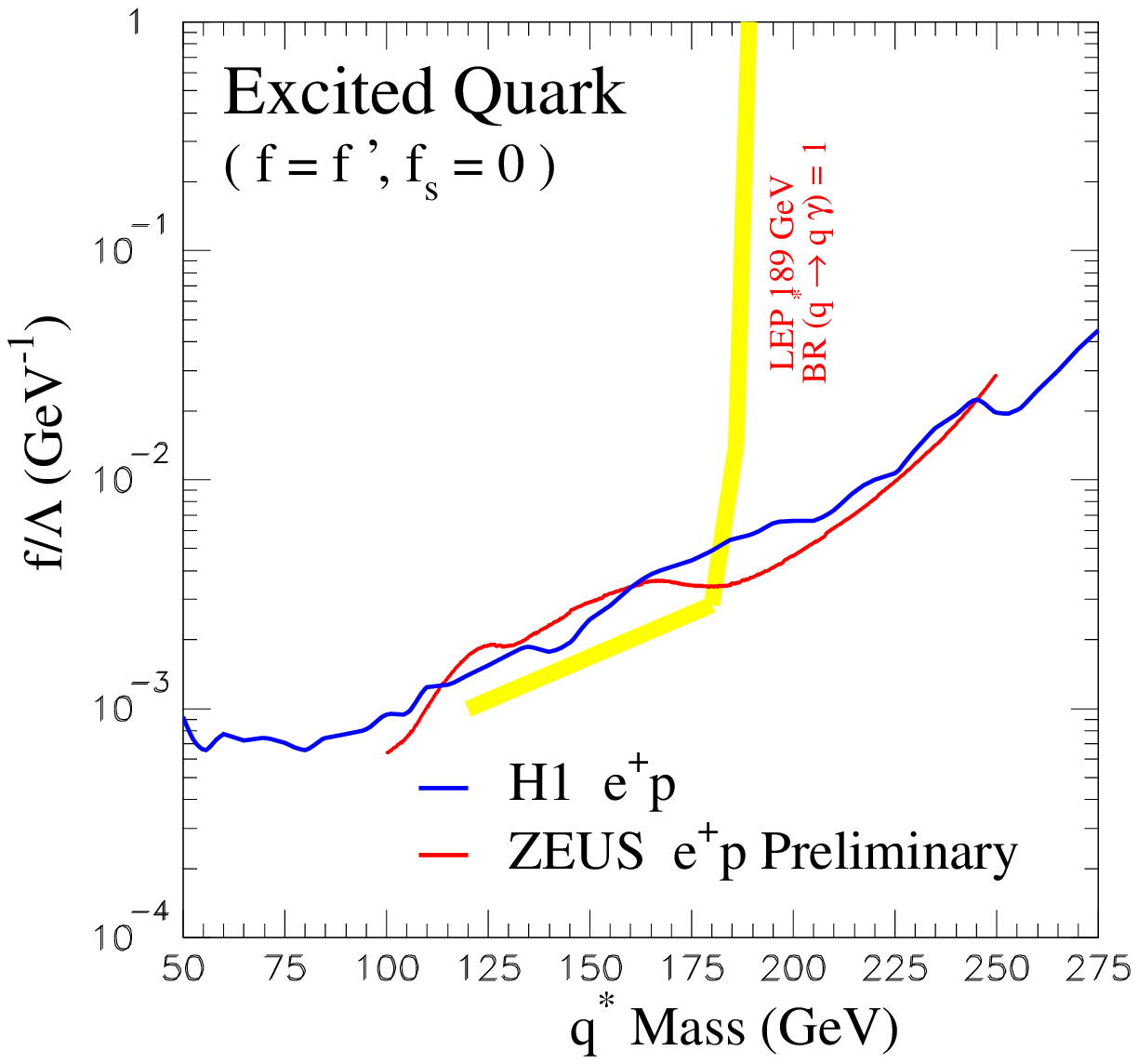,width=3.7cm}&
\hspace*{-1.4cm}\epsfig{file=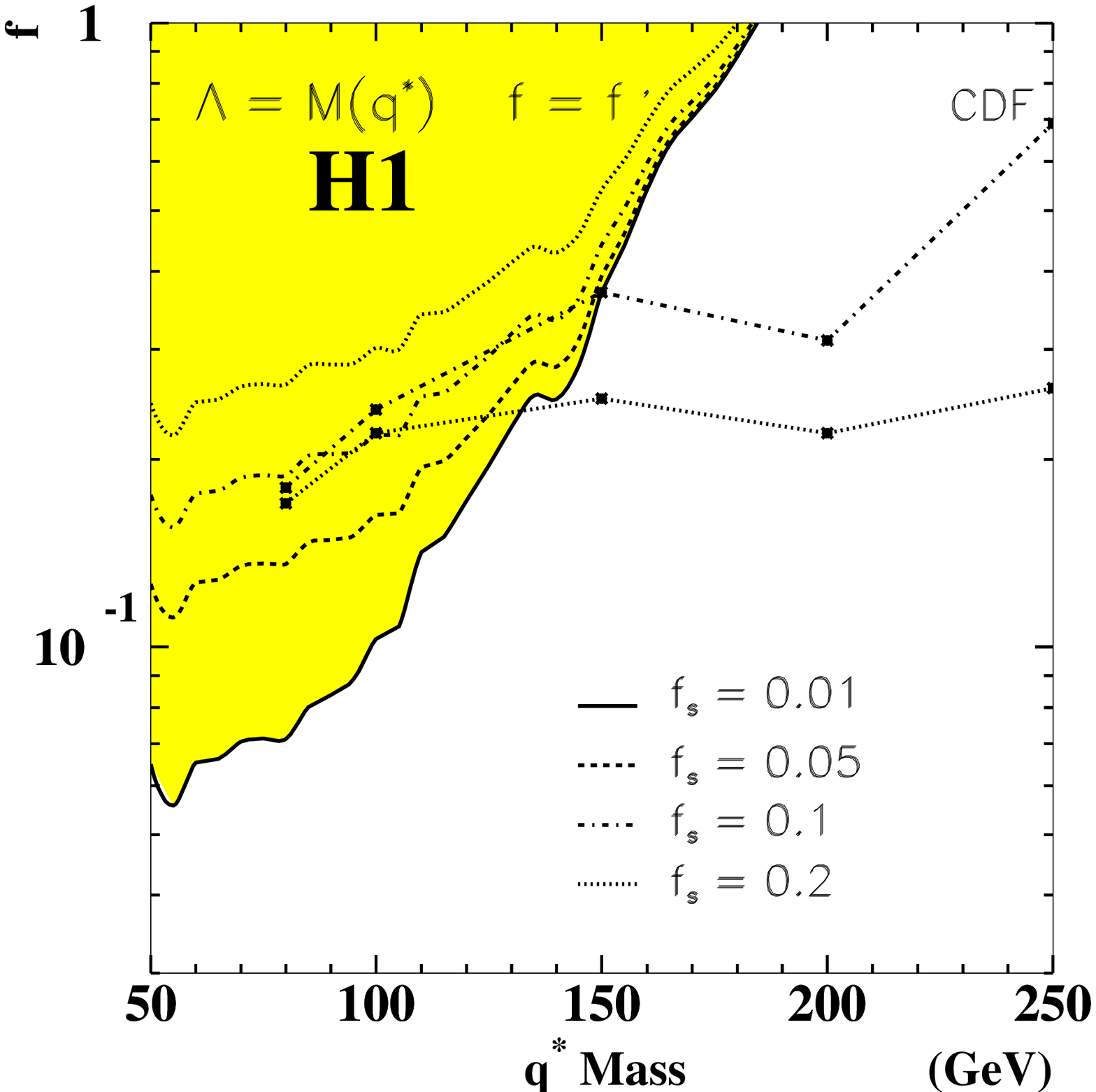,width=4.cm}
\end{tabular}
\end{center}
\caption{{\it Left:} Limits on the coupling constant as a function of the mass of the excited electron. {\it Center:} Limit on the coupling constant as a function of the mass of the excited quark. {\it Right:} Limit on the form factor $f$ as a function of the mass of the excited quark, with the hypothesis $f=f'$ and $\Lambda = M(q^*)$ for various values of the form factor $f_s$.\label{fig:EXq}}
\vspace*{-.5cm}
\end{figure}

The search in the data collected at $\sqrt{s} = 300 $ GeV between 1994 and 1997 has been completed~\cite{Adloff:2000gv,osk-1040} and no signal was found, thus limits have been inferred (see figure~\ref{fig:EXq}). By assuming $\frac{f}{\Lambda} = \frac{1}{M^*}$ one can exclude excited electrons of mass $M^*$ lower than 229 GeV and excited quarks of mass lower than 203 GeV.

\begin{figure}[tbp]
\vspace*{-.6cm}
\begin{center}
\begin{tabular}{cc}
\epsfig{file=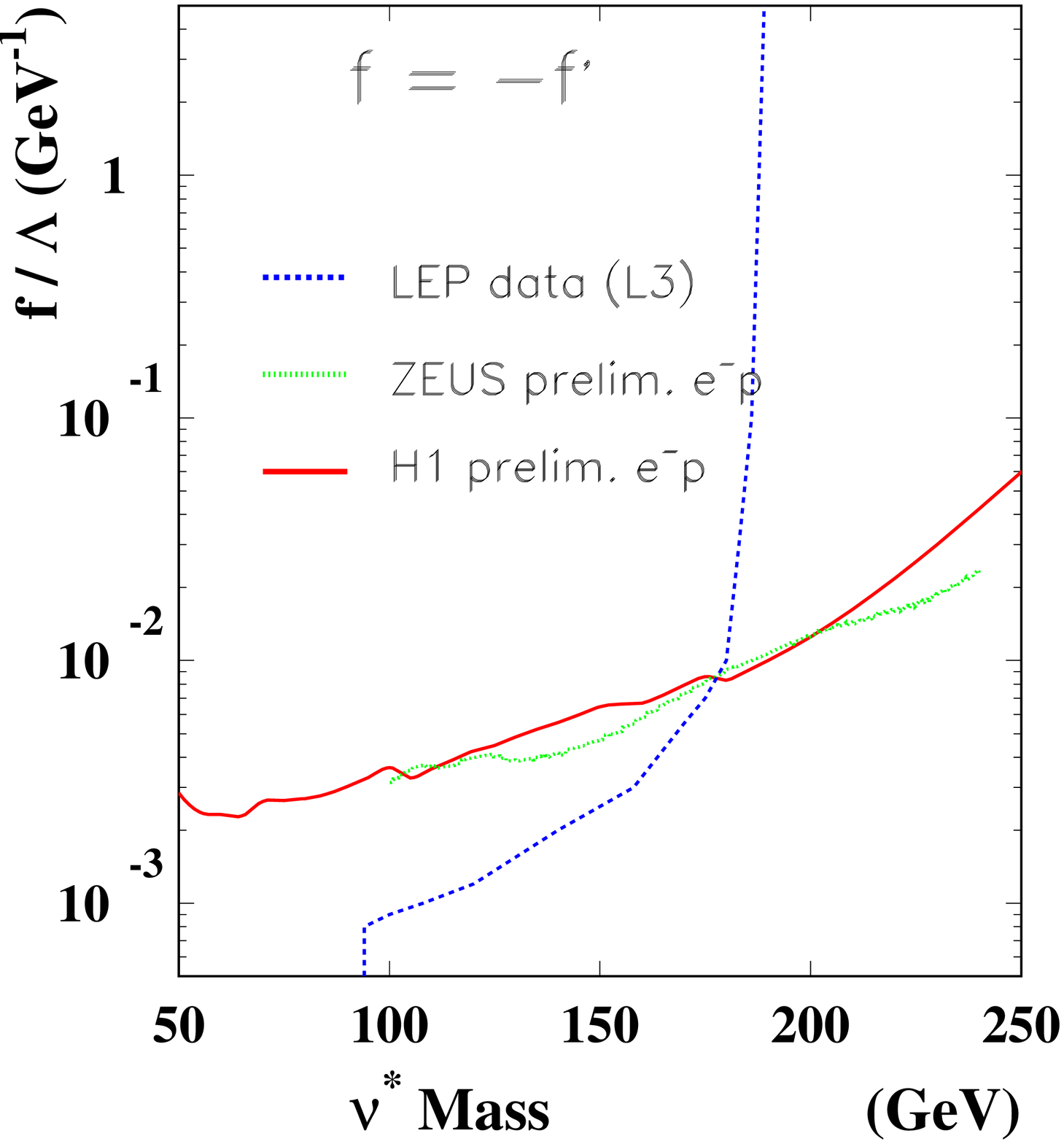,width=3.5cm}&
\epsfig{file=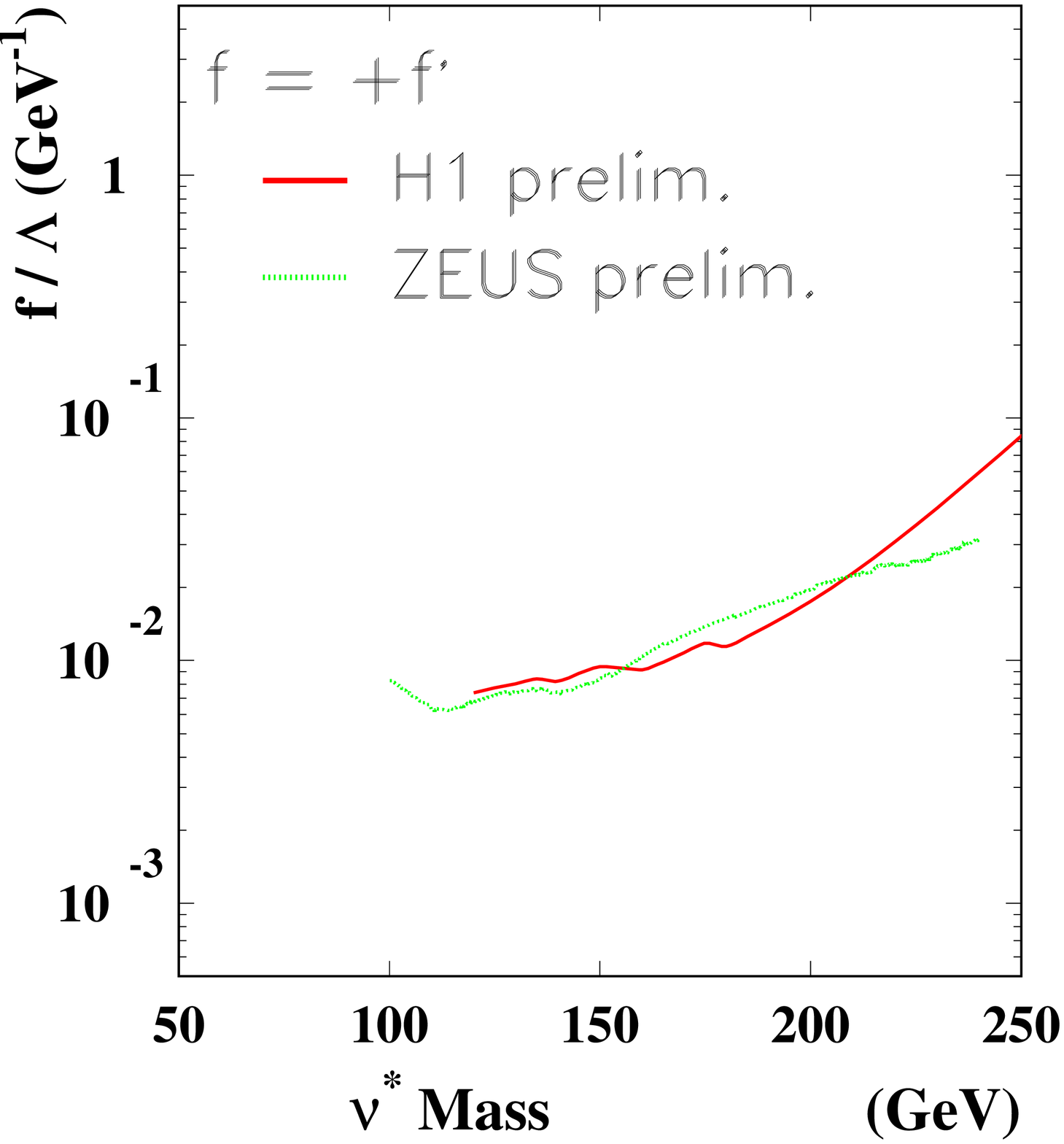,width=3.5cm}
\end{tabular}
\end{center}
\vspace*{-.7cm}
\caption{Limits on the coupling constant as a function of the mass of the excited neutrino for the hypothesis $f=-f'$ (left) and $f=+f'$ (right). \label{fig:EXnu}}
\vspace*{-.5cm}
\end{figure}

The production cross-section for excited neutrinos ($\nu^*$) is higher (by two orders of magnitude for a $\nu^*$ of 200 GeV) in $e^-p$ data than in $e^+p$ data. Preliminary results show that no evidence for $\nu^*$ production has been found by either collaboration~\cite{osk-1040,osk-956} in the recently taken $e^-p$ data. Limits have thus been inferred (see figure~\ref{fig:EXnu}). 
By assuming $\frac{f}{\Lambda} = \frac{1}{M^*}$ one can exclude excited neutrinos  of mass below 150 GeV for $f=-f'$ and below 134 GeV for $f=+f'$. As one can see on the figures~\ref{fig:EXq} (center) and \ref{fig:EXnu}, the limits set at HERA reach higher energies than those set at LEP. On figure~\ref{fig:EXq}, the plot on the right shows that for small $f_s$ HERA can extend the limits set at the TeVatron on the excited quark coupling.

\section{Events with an isolated lepton and missing transverse momentum, FCNC}

\begin{figure}[htbp]
\vspace*{-2.cm}
\begin{center}
\epsfig{file=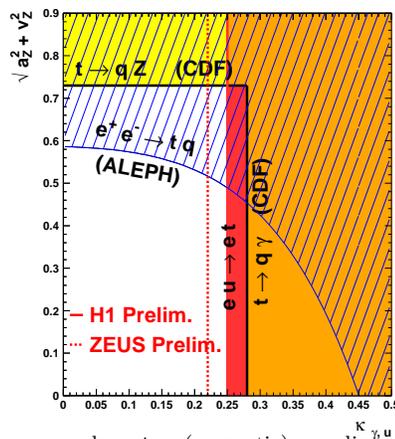,width=6.cm}
\end{center}
\caption{Limits on the anomalous $tu\gamma$ (magnetic) coupling $\kappa_{\gamma,u}$ and on the anomalous $tuZ$ (vector) coupling $\sqrt{a^2_Z+v^2_Z}$ obtained at the TeVatron (CDF), LEP (ALEPH) and HERA (H1 and ZEUS) colliders.\label{fig:top}}
\vspace*{-.5cm}
\end{figure}

Puzzling events with high $P_t$ isolated lepton and missing transverse momentum have been previously observed at HERA. In the SM, such final states mainly come from W production. Both collaborations 
have a comparable analysis~\cite{osk-974,osk-1041} (see table~\ref{tab:isol}). H1 still sees a deviation from the SM prediction whereas ZEUS does not and this needs to be resolved with more statistics. Beyond the SM, an explanation of these events could be flavour changing neutral currents (FCNC) leading to single top production which is highly suppressed in the SM. 
Limits on this phenomena have been set by both collaborations~\cite{osk-1041,osk-961} (see figure \ref{fig:top}).

\begin{table}
\begin{center}
\begin{tabular}{|c||c|c|c|c||}
\hline
          &   H1 data & \multicolumn{2}{|c|}{SM expectation } & ZEUS data \\
 & preliminary & H1 & ZEUS & preliminary \\
\hline
$P^X_T (e)  > 25 $ GeV & 3  & $ 0.84 $ & $ 0.78 $ & 1 \\ 
\hline
$P^X_T (\mu)  > 25 $ GeV & 6  &\hspace*{.1cm} $ 0.94 $\hspace*{.1cm} &\hspace*{.1cm} $ 0.82 $\hspace*{.1cm} & 0 \\ 
\hline
\end{tabular}
\end{center}
\caption{Comparison of the numbers of  events with high $P_t$ isolated leptons compared to the Standard Model expectation. The results presented here correspond, for both collaborations, to $ 82 pb^{-1}$ analysed before July 2000. In the H1 numbers, further cuts were applied after the nominal selection cuts such that both analyses correspond to similar acceptances for W events.\label{tab:isol}}
\vspace*{-.5cm}
\end{table}

\section{Conclusion}

Searches have been performed for contact interactions, extra dimensions, leptoquarks, \rp supersymmetry and excited fermions and no evidence was found for any of these processes. Limits have thus been derived and are comparable to those set at LEP or the TeVatron. The isolated lepton events observed by H1 are still puzzling. HERA is currently being upgraded to produce 5 times more luminosity starting this autumn (2001), thus searches have an exciting future at HERA.

\section*{Acknowledgments}

\vspace*{-.2cm}

I wish to thank all my ZEUS and H1 colleagues who have contributed to the results discussed here.

\end{document}